\documentclass[aps,prb,preprint,groupedaddress,showpacs,amssymb]{revtex4}
\usepackage{graphicx}
\usepackage{dcolumn}
\usepackage{bm}

\begin{document}

\preprint{Submitted to Phys. Rev. B}

\title{Anomalous $f$-electron Hall Effect in the Heavy-Fermion System CeTIn$_{5}$ (T = Co, Ir, or Rh)}

\author{M.F. Hundley}
\author{A. Malinowski}
\author{P.G. Pagliuso}
 \altaffiliation[Present address: ]{Instituto de F\'{i}sica ``Gleb Wataghin,'' UNICAMP, 13083-970 Campinas, Brazil.}
\author{J.L. Sarrao}
\author{J.D. Thompson}
\affiliation{Materials Science and Technology Division, Los Alamos
National Laboratory, Los Alamos, NM 87545}

\date{\today}

\begin{abstract}

The in-plane Hall coefficient $R_{H}(T)$ of CeRhIn$_{5}$,
CeIrIn$_{5}$, and CeCoIn$_{5}$ and their respective non-magnetic
lanthanum analogs are reported in fields to 90 kOe and at
temperatures from 2 K to 325 K. $R_{H}(T)$ is negative,
field-independent, and dominated by skew-scattering above $\sim$
50 K in the Ce compounds. $R_{H}(H \rightarrow 0)$ becomes
increasingly negative below 50 K and varies with temperature in a
manner that is inconsistent with skew scattering. Field-dependent
measurements show that the low-T anomaly is strongly suppressed
when the applied field is increased to 90 kOe. Measurements on
LaRhIn$_{5}$, LaIrIn$_{5}$, and LaCoIn$_{5}$ indicate that the
same anomalous temperature dependence is present in the Hall
coefficient of these non-magnetic analogs, albeit with a reduced
amplitude and no field dependence. Hall angle ($\theta_{H}$)
measurements find that the ratio
$\rho_{xx}/\rho_{xy}=\cot(\theta_{H})$ varies as $T^{2}$ below 20
K for all three Ce-115 compounds. The Hall angle of the La-115
compounds follow this T-dependence as well. These data suggest
that the electronic-structure contribution dominates the Hall
effect in the 115 compounds, with $f$-electron and Kondo
interactions acting to magnify the influence of the underlying
complex band structure. This is in stark contrast to the situation
in most $4f$ and $5f$ heavy-fermion compounds where the normal
carrier contribution to the Hall effect provides only a small,
T-independent background to $R_{H}.$
\end{abstract}


\maketitle

\section{\label{intro}Introduction}
The discovery of $f$-electron compounds that display a unique
combination of competing ground states promotes advances in the
field of Kondo physics by challenging our understanding of the
underlying many-body interactions.\cite{reviewHF} The flurry of
research activity associated with the Ce$_{n}$T$_{m}$In$_{3n+2m}$
(T = Co, Ir, or Rh; $n$ = 1 or 2; $m$ = 0 or 1) compounds
certainly fits this description.\cite{Thompson01,Thompson03} The
compounds that reside within this ``family'' exhibit essentially
all of the many ground states that have been observed in
$f$-electron systems, including paramagnetism, antiferromagnetism,
and exotic ambient-pressure and pressure-induced
superconductivity. These compounds also exhibit a mixture of
essentially all known $f$-electron phenomena, including a
Kondo-renormalized ground state, Fermi-liquid behavior, and both
pressure-induced and ambient-pressure non-fermi-liquid (NFL)
behavior in the vicinity of a antiferromagnetic quantum-critical
point (QCP).\cite{Thompson03} To date, our understanding of
$f$-electron Kondo systems has rested, in part, on the competition
between Kondo and Ruderman-Kittel-Kasuya-Yosida (RKKY)
interactions.\cite{Doniach77} The complex phase diagram defined by
the Ce$_{n}$T$_{m}$In$_{3n+2m}$ family challenges this
understanding.

The Ce$_{n}$T$_{m}$In$_{3n+2m}$ family contains three distinct
subgroups, each with  differing degrees of anisotropy. The first
subgroup ($m$ = 0) contains a single member, the cubic
antiferromagnet (AFM) CeIn$_{3}$. At ambient pressure CeIn$_{3}$
orders magnetically at $T_{N}$ = 10 K and has a slightly enhanced
Sommerfeld coefficient $\gamma$ ($\equiv C_{p}/T$ as $T
\rightarrow 0)$ of 100 mJ/mole K$^{2}$.\cite{Lawrence81} An
applied pressure of $P_{c}\approx$ 25 kbar drives $T_{N}$ to
zero,\cite{Walker97,Mathur98} leads to NFL
behavior,\cite{Steiner01} and produces a superconducting state
with a maximum transition temperature $T_{c}$ = 0.2 K. The second
subgroup ($n$ = $m$ = 1) contains three known Ce-based members,
CeCoIn$_{5}$, CeIrIn$_{5}$, and CeRhIn$_{5}$. These tetragonal
Ce-115 compounds have a quasi-two-dimensional structure that is
composed of a cubic CeIn$_{3}$ element separated by a TIn$_{2}$
layer. CeRhIn$_{5}$ is an ambient pressure antiferromagnet
($T_{N}$ = 3.8 K) with an enhanced $\gamma$ $\approx$ 400 mJ/mole
K$^{2}$.\cite{Hegger00} Superconductivity occurs at $T_{c}$ = 2.1
K in a pressure of 16 kbar.\cite{Hegger00} In contrast
ambient-pressure
unconventional\cite{Movshovich01,Izawa01,Kohori01}
superconductivity occurs in both CeIrIn$_{5}$ (T$_{c}$ = 0.4
K)\cite{Petrovic01a} and CeCoIn$_{5}$ ($T_{c}$ = 2.3
K).\cite{Petrovic01b} In both superconductors many-body
interactions produce a highly renormalized mass state
($\gamma$(CeIrIn$_{5}$) = 0.75 J/mole K$^{2}$,
$\gamma$(CeCoIn$_{5}$) $\approx$ 1 J/mole K$^{2}$) that becomes
evident in the specific heat $C_{p}$(T) below 10 K. CeCoIn$_{5}$
exhibits clear NFL behavior in resistivity $\rho$(T) and
$C_{p}$(T) data, indicating that this compound resides near a
antiferromagnetic quantum-critical point.\cite{Sidorov02,Kohori01}
While the specific-heat of CeIrIn$_{5}$ exhibits fermi-liquid
behavior just above $T_{c}$,
 $\rho(T)$, thermal-expansion, and $1/T_{1}$ data suggest that this compound may be close to a
QCP as well.\cite{Kohori01,Oeschler03} The third subgroup ($n$ =
2, $m$ = 1) is composed of three double-layer members
Ce$_{2}$TIn$_{8}$ (T = Rh, Ir, Co). These Ce-218 compounds exhibit
the same phenomena seen in the Ce-115 materials, including
paramagnetic\cite{Moreno02} and AFM\cite{Cornelius01}
ground-states, pressure-induced and ambient pressure
superconductivity,\cite{Nicklas03,Chen02} and NFL
behavior.\cite{Chen03}

Anisotropy is a feature common to nearly all of the properties
exhibited by the Ce-115 compounds. In part, these anisotropies
stem from the tetragonal 115 lattice structure. Band structure
calculations and de Haas-van Alphen (dHvA) measurements indicate
that the 115 materials have a complicated quasi-2D Fermi surface
(FS) composed of multiple electron and hole
orbits.\cite{Cornelius00,Haga01,Hall01a,Hall01b,Shishido02} When
considering the novel properties of the 115 compounds it is
critical to differentiate between phenomena that are controlled by
prosaic single-electron physics and those determined by correlated
electron interactions.

The Hall effect provides a useful means of elucidating the
relative importance of single-electron (i.e., conventional
electronic structure) and many-body interactions in $f$-electron
systems. The Hall effect in these systems is strongly influenced
by the scattering of charge carriers via the orbital angular
momenta of localized electrons.\cite{Coleman85,Hadzic86,Fert87} In
most $f$-electron compounds this orbital skew scattering effect
dominates the Hall coefficient $R_{H}$(T) at temperatures greater
than the characteristic Kondo temperature $T_{K}$. In contrast,
the electronic-structure component is generally a small,
temperature-independent contributor to $R_{H}$(T). Many body
correlations frequently dominate the Hall coefficient as the
system approaches the Kondo temperature from
above.\cite{Fert87,CohHall} At and below $T_{K}$, $R_{H}$(T) is
influenced by the onset of coherence and the development of the
Abrikosov-Suhl resonance in the electronic density of states near
the Fermi energy.\cite{Winzer86,Huth94} For these reasons T and
H-dependent Hall measurements provide insights into the relative
importance of electronic structure and many-body interactions in
the physical properties of $f$-electron compounds.

We have measured the in-plane Hall coefficients of CeRhIn$_{5}$,
CeIrIn$_{5}$, and CeCoIn$_{5}$ and their respective non-magnetic
lanthanum analogs in fields to 90 kOe and at temperatures from 2 K
to 325 K. The Hall effect is dominated by skew-scattering above
$\sim$ 50 K in the Ce-115 compounds, and a precipitous negative
drop is present  in the in-plane Hall coefficient below 50 K that
is inconsistent with incoherent orbital scattering. This same
temperature dependence, but with a more modest T-dependent
amplitude, is evident in Hall measurements on the non-magnetic
analogs LaRhIn$_{5}$, LaIrIn$_{5}$, and LaCoIn$_{5}$ signifying
that the Hall effect in the Ce-115 materials is dominated by the
conventional Hall carrier contribution. Measurements on the Ce-115
compounds indicate that the Hall anomaly present below 50 K can be
suppressed significantly by a 90 kOe field, suggesting that
field-dependent many-body Kondo interactions influence the unusual
Hall effect intrinsic to the 115 electronic structure. Lastly,
Hall angle ($\theta_{H}$) measurements indicate that the ratio
$\rho_{xx}/\rho_{xy} = \cot(\theta_{H}$) varies as $T^{2}$ below
20 K in all three Ce-115 compounds; La-115 $\cot(\theta_{H}$) data
between 30 K and 100 K vary quadratically with temperature too.
This Hall-angle temperature dependence has been observed in
high-T$_{c}$ cuprates\cite{Chien91} as well, and some have
speculated that this behavior is linked to antiferromagnetic spin
fluctuations due to a nearby QCP.\cite{Kontani99,Nakajima03} While
it is tempting to make the same connection for these Ce
heavy-fermion compounds, the fact that $\cot(\theta_{H})\approx
T^{2}$ in all six 115 compounds suggests that this temperature
dependence is unconnected with QCP-related spin-fluctuations.

\section{\label{exp details}Experimental Details}

Single crystal Ce-115 and La-115 transport samples were produced
with an indium flux-growth technique. X-ray diffraction on
powdered crystals indicates that the crystals are single-phase and
form in the primitive tetragonal HoCoGa$_{5}$ structure. While the
excess In flux is critical for growing the crystals, residual
indium has the negative side effect of contaminating essentially
all transport measurement on as-grown samples. To eliminate this
problem all samples employed in this study were polished and then
pre-screened via $\rho$(T) and magnetic susceptibility $\chi$(T)
measurements to ensure that no extrinsic (In) superconductivity
was evident at 3.4 K.

Hall measurements were made on single-crystal samples that were
cut and polished into thin Hall bars with sample thicknesses
ranging from 0.1 to 0.4 mm. All Hall coefficients reported here
are in-plane measurements with the magnetic field applied along
the c axis. The standard four-terminal contact arrangement was
used, and the contacts were made with silver conductive epoxy. The
longitudinal resistivity ($\rho_{xx}$) and the Hall voltage
V$_{H}$ were measured with a low-frequency resistance bridge.
Systematic errors associated with Hall contact misalignment were
eliminated by ascribing the asymmetric component of the
field-reversed transverse voltage to the Hall voltage, $V_{H}(H) =
[V_{H}(+H) - V_{H}(-H)]/2$.\cite{Wieder79} The Hall coefficient
was calculated from the standard expression R$_{H}=V_{H}t/IH$,
where $I$ is the applied current flowing perpendicular to the
applied field $H$, and $t$ is the sample thickness.

The Hall measurements were made in fields from 1 kOe to 90 kOe in
order to examine the $R_{H}$ field dependence and to determine the
Hall coefficient in the low-field limit, $R_{H}(H \rightarrow 0)$.
The Hall coefficient is extremely field-dependent in the Ce-115
compounds below 20 K; this strong field-dependence coupled with
the fact that $V_{H}$ = 0 in the zero field limit means that some
care is required to accurately determine $R_{H}(H \rightarrow 0)$.
We employed two methods in determining R$_{H}(H \rightarrow 0)$
from the measured Hall voltages. In method one $R_{H}(H)$ was
determined from the measured $V_{H}(H)$ in fields ranging from 1
to 90 kOe, and $R_{H}(H \rightarrow 0)$ was calculated by
extrapolating the data to $H$ = 0. In method two $R_{H}(H
\rightarrow 0)$ was determined from the zero-field limit of the
Hall resistivity field derivative, $R_{H}(0)$ =
$\partial\rho_{xy}/\partial H$, where $\rho_{xy} = V_{H}t/I$. Both
methods produce $R_{H}(H \rightarrow 0)$ values that are identical
within experimental error. In contrast to the strongly
field-dependent Hall voltage in the Ce-115 compounds, the La-115
materials exhibit a field-independent Hall coefficient (i.e.,
$\rho_{xy} \propto H$). As such, the T-dependent La-115 Hall data
reported in Sec. \ref{results} were measured in 10 kOe.

\section{\label{results}Results}

\subsection{\label{La115}Properties of LaTIn$_{5}$ (T = Co, Ir, and Rh)}

\begin{figure}
 \includegraphics{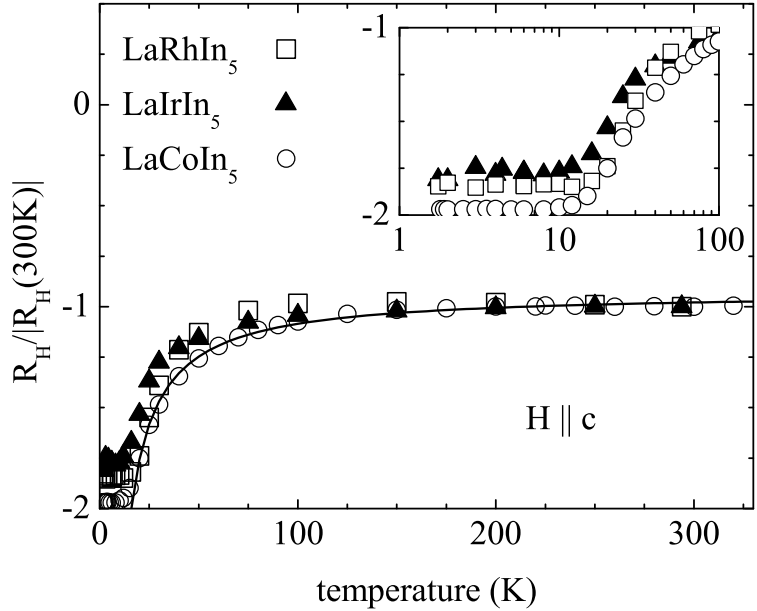}
 \caption{\label{fig:HallLa}In-plane Hall coefficients (H = 10 kOe) plotted as a function of temperature for LaRhIn$_{5}$, LaIrIn$_{5}$,
 and LaCoIn$_{5}$. Low-temperature data are highlighted in the inset.
 The data are normalized by the absolute value of each compound's 300 K Hall coefficient. The solid line is a fit to the
 LaCoIn$_{5}$ data utilizing Eq. (\ref{eqn:LaHall}).}
\end{figure}

The La-115 compounds are isostructural analogs of the Ce-115
compounds. As such, the La-115 transport properties are indicative
of the non-magnetic contributions to transport in the Ce
compounds. The characteristics of these La-115 compounds are, for
the most part, unremarkable and typical of a non-magnetic
intermetallic system. LaRhIn$_{5}$, LaIrIn$_{5}$, and LaCoIn$_{5}$
exhibit a temperature-independent Pauli paramagnetic
susceptibility, and the resistivity  varies linearly with
temperature above $\sim$ 50 K. Room-temperature resistivities
range from 10 to 20 $\mu\Omega$ cm, while anisotropic resistivity
measurements on LaRhIn$_{5}$ find that the nonmagnetic electronic
anisotropy inherent in the tetragonal 115 structure is relatively
small.\cite{Christianson02} The low-temperature (10 K)
magnetoresistance (MR) of LaRhIn$_{5}$ and LaIrIn$_{5}$ is
positive and grows quadratically with field. These properties are
consistent with a simple metallic system.

The temperature-dependent in-plane Hall coefficients of the three
La-115 compounds are displayed in Fig. \ref{fig:HallLa}. All three
La-115 compounds exhibit a negative Hall coefficient, with 300 K
values of -7.5 x 10$^{-10}$ m$^{3}$/C, -5.1 x 10$^{-10}$
m$^{3}$/C, and -4.2 x 10$^{-10}$ m$^{3}$/C for the Co, Rh, and Ir
compounds, respectively. R$_{H}$ is nearly temperature-independent
above 100 K, drops monotonically below 100 K, and begins to
saturate below 20 K. The inset to Fig. \ref{fig:HallLa} shows that
$R_{H}$ is essentially T-independent below 10 K for the three
La-115 compounds. While the data displayed in Fig.
\ref{fig:HallLa} were measured in a 10 kOe field, measurements
made in fields from 1 to 90 kOe show that R$_{H}$ is
field-independent ($\rho_{xy}\propto H$) at 5, 10, 50, 100, and
300 K. The data in Fig. \ref{fig:HallLa} are normalized by the
absolute value of each compound's 300 K Hall coefficient to
underscore the fact that $R_{H}$ follows essentially the same
temperature-dependence in all three La-115 compounds. The data
above 20 K are well described by the expression
\begin{equation}
R_H^{(La)}  = R_H^\infty  + \frac{1}{{a + bT}} \label{eqn:LaHall},
\end{equation}
where $R_H^\infty$, $a$, and $b$ are fitting parameters. Eq.
(\ref{eqn:LaHall}) was used to produce the fit to the normalized
LaCoIn$_{5}$ data that is shown in Fig. \ref{fig:HallLa}; the
fitting parameters are $a$ = 0.07, $b$ = -0.06 K$^{-1}$, and
$R_H^\infty$ = -0.93. Surprisingly, the same expression (with
$R_H^\infty$ = 0) describes the normal-state Hall response in both
YBa$_{2}$Cu$_{3}$O$_{7}$ and
MgB$_{2}$.\cite{Chien91,Ginsberg93,Jin01} The substantial
T-dependence present below 100 K in the La-115 data is quite
unusual since metals typically have a nearly constant Hall
coefficient. A T-dependent Hall coefficient is usually a sign that
the underlying electronic structure is composed of multiple
electron and hole bands with differing mobility temperature
dependencies. Band-structure calculations and dHvA measurements do
find that the La-115 electronic structure is very
complex,\cite{Cornelius00,Haga01,Hall01a,Hall01b,Shishido02} so
the La-115 $R_{H}$ temperature dependence appears to be a
reflection of the 115 electronic structure.

\begin{figure}
 \includegraphics{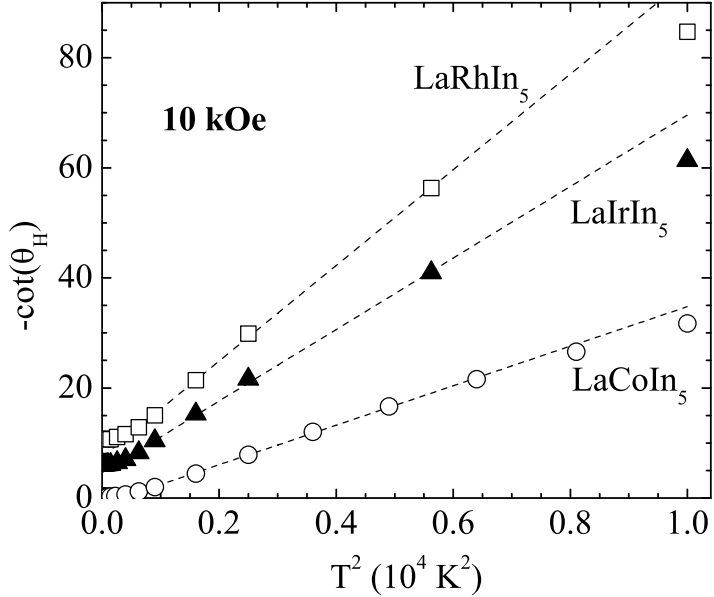}
 \caption{\label{fig:cotLa}The in-plane Hall angle $\cot(\theta_{H}$) plotted as a function of $T^{2}$ for LaRhIn$_{5}$,
 LaIrIn$_{5}$, and LaCoIn$_{5}$. The dashed lines are
 linear fits to the data. For clarity the Ir and Rh data have been vertically offset by 5 and 10, respectively.}
\end{figure}

We next consider the La-115 Hall angle ($\theta_{H}$), which is
plotted as $-\cot(\theta_{H})$ vs $T^{2}$ in Fig. \ref{fig:cotLa}.
$\theta_{H}$ is defined as the angle between the applied current
and the resulting electric field, and it is determined
experimentally via $\cot(\theta_{H})$ = $\rho_{xx}/\rho_{xy} =
\rho_{xx}/R_{H}H$. The data fall on a straight line from roughly
30 to 100 K for all samples, indicating that $\cot(\theta_{H})$
varies with temperature as
\begin{equation}
\cot (\theta _H ) = \alpha  + \beta T^2 \label{eqn:LaCot}.
\end{equation}
This quadratic temperature dependence occurs over precisely the
same temperature range where $R_{H}$ is extremely temperature
dependent. The $T^{2}$ behavior results from the combination of a
resistivity that varies linearly with $T$ and a Hall coefficient
that varies inversely with $T$. The Hall angle of both
YBa$_{2}$Cu$_{3}$O$_{7}$ [\onlinecite{Chien91,Ginsberg93}] and
MgB$_{2}$ [\onlinecite{Jin01}] follow the same anomalous
temperature dependence. The La-115 compounds exhibit a
conventional Hall angle temperature dependence ($\cot(\theta_{H})
\sim T$) only above 100 K where $R_{H}$ is nearly constant.

\subsection{\label{Ce115}Properties of CeTIn$_{5}$ (T = Co, Ir, and Rh)}

\begin{figure}
 \includegraphics{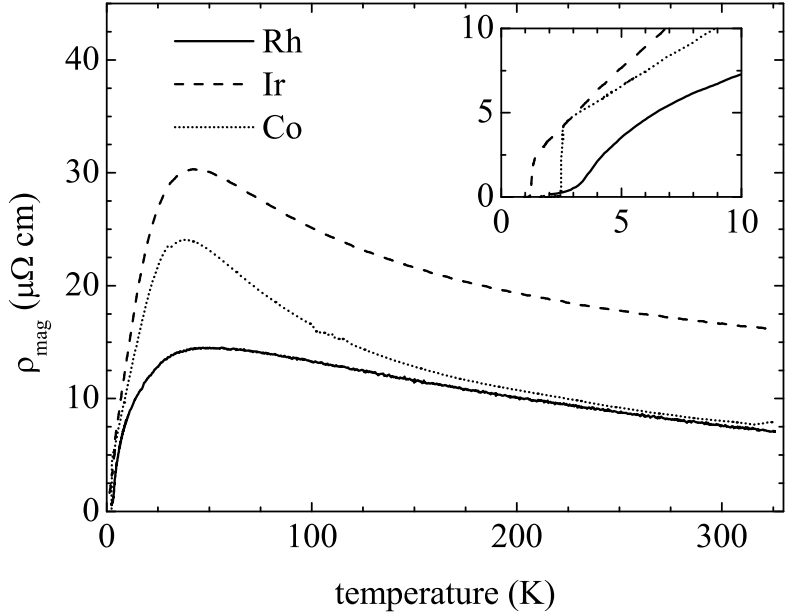}
 \caption{\label{fig:rhomag}Magnetic-scattering contributions to the
 in-plane resistivity plotted as a function of temperature for CeRhIn$_{5}$, CeIrIn$_{5}$, and CeCoIn$_{5}$;
 data below 10 K are highlighted in the inset.}
\end{figure}

Figure \ref{fig:rhomag} shows the magnetic resistivity
$\rho_{mag}$ of CeRhIn$_{5}$, CeIrIn$_{5}$, and CeCoIn$_{5}$
plotted as a function of temperature from 1 to 325 K. The
resistivity contribution from magnetic scattering is calculated by
subtracting the electron-phonon contribution (the resistivity of
the non-magnetic La analog) from the Ce-115 resistivity,
$\rho_{mag} = \rho_{Ce} - \rho_{La}$. $\rho_{mag}(T)$ exhibits a
broad maximum located below 50 K in all three compounds; the
coherence temperature $T_{coh}$ is defined by the temperature
where the resistivity peaks. Although the resistivity maximum in
the CeRhIn$_{5}$ data is not as pronounced as those in the
CeIrIn$_{5}$ and CeCoIn$_{5}$ data, $\rho_{mag}$ varies roughly as
$-\ln(T)$ in each of the Ce-115 materials for $T > T_{max}$. The
inset to Fig. \ref{fig:rhomag} shows the low-temperature behavior
of the magnetic resistivity. The superconducting transitions in
CeIrIn$_{5}$ and CeCoIn$_{5}$ are clearly evident as abrupt drops
in $\rho_{mag}$ at their respective transport $T_{c}$'s, while the
onset of magnetic order at $T_{N}$ (3.8 K) in CeRhIn$_{5}$ leads
to a more subtle inflection-point anomaly. Just above their
superconductivity transitions, $\rho(T)$ for CeIrIn$_{5}$ and
CeCoIn$_{5}$ varies with temperature as $T^{1.3}$ and $T^{1.0}$,
respectively.\cite{Petrovic01a,Sidorov02} These power-laws differ
from that of a Fermi-liquid ($\rho \sim T^{2}$), and are
suggestive of non-Fermi-liquid behavior.

\begin{figure}
 \includegraphics{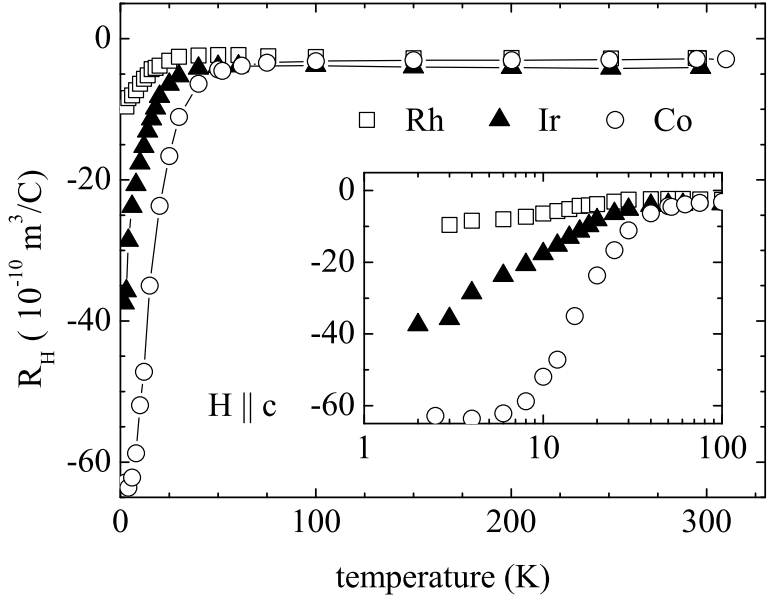}
 \caption{\label{fig:HallCe}Low-field ($H \rightarrow 0$) in-plane ($H \parallel$ c) Hall coefficients
 plotted as a function of temperature for CeRhIn$_{5}$ ($\square$), CeIrIn$_{5}$ ($\blacktriangle$), and CeCoIn$_{5}$ ($\circ$).
 Low-temperature data are highlighted in the inset.}
\end{figure}

\begin{figure}
 \includegraphics[width=0.51\textwidth, trim= 18 10 0 15]{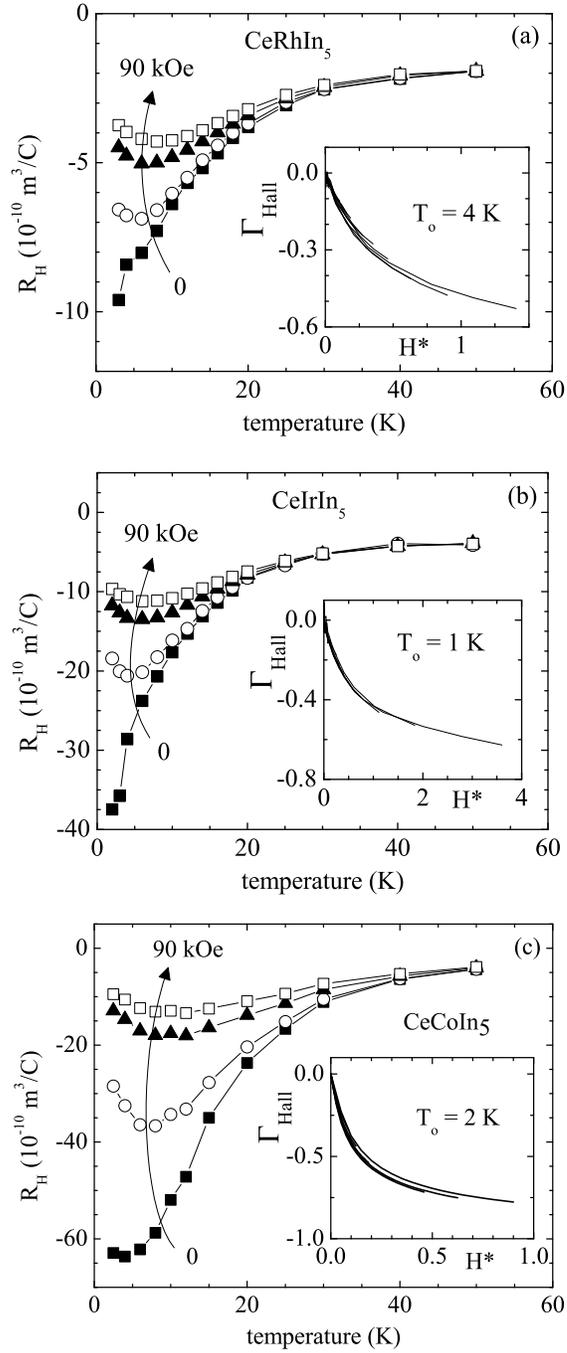}
 \caption{\label{fig:HallH}$R_{H}(T)$ measured in various applied fields plotted as a function of temperature
 for CeRhIn$_{5}$ (a), CeIrIn$_{5}$ (b), and CeCoIn$_{5}$ (c).
 Data are shown at four fields: $H \rightarrow$
 0 ($\blacksquare$), 10 kOe ($\circ$), 50 kOe ($\blacktriangle$),
 and 90 kOe ($\square$). The insets show isotherms of the relative change in the field-dependent Hall coefficient $\Gamma_{Hall}$ plotted as
 a function of the scaled field $H^{*}$ with $\beta$ = 2 (see text); the data were measured at fixed temperatures between 4 and 30
 K, and the $T_{o}$ parameters used to determine $H^{*}$ are listed in each inset.}
\end{figure}

Figure \ref{fig:HallCe} shows the low-field ($H \rightarrow$ 0)
Hall coefficients of CeRhIn$_{5}$, CeIrIn$_{5}$, and CeCoIn$_{5}$
plotted as a function of temperature between 2 and 300 K; the
low-temperature behavior is highlighted in the figure's inset.
$R_{H}(T)$ is electron-like at all temperatures, and the
precipitous drop that occurs in the Hall response of all three
materials below 50 K is certainly the most prominent feature in
the data. The data displayed in the inset indicate that the Hall
response saturates below 4 K, and that the drop in $R_{H}(T)$
occurs at roughly the same temperature ($\sim$40 K) in the three
compounds. While the three Ce-115 compounds all show signs of the
same drop in $R_{H}$ below 40 K, the feature is far more prominent
for CeCoIn$_{5}$ and CeIrIn$_{5}$ than for CeRhIn$_{5}$. The Hall
voltage in CeCoIn$_{5}$ drops to zero below 2.3 K for fields less
than $H_{c2}$ due to the onset of superconductivity (this data is
omitted for clarity in Fig. \ref{fig:HallCe}). With regard to the
AFM transition in CeRhIn$_{5}$, no discernable anomaly is present
in $R_{H}$ at or below $T_{N}$. At higher temperatures $R_{H}$ is
weakly temperature dependent in all three Ce-115 compounds. The
Hall coefficients of CeRhIn$_{5}$ and CeIrIn$_{5}$ have a positive
slope above 50 K; between 50 K and 300 K $|R_{H}|$ grows by 25\%
for CeRhIn$_{5}$, and 9\% for CeIrIn$_{5}$. The CeCoIn$_{5}$ Hall
coefficient shows no such positive slope above 100 K; instead, the
CeCoIn$_{5}$ Hall response drops by roughly 10\% between 100 K and
300 K. The 300 K Hall coefficient is nearly the same in the three
Ce-115 compounds, and the room temperature value (-3.5 x
10$^{-10}$ m$^{3}$/C) corresponds to an \emph{effective} carrier
concentration of 2.9 $e^{-}$ per formula unit.

Figure \ref{fig:HallH} displays the Ce-115 Hall coefficients for
temperatures below 60 K when measured in four different fields ($H
\rightarrow$ 0, 10, 50, and 90 kOe). Increasing field strength
qualitatively influences the Hall response in the three Ce-115
compounds in the same manner. The data indicate that the Hall
response is extremely field-dependent in the temperature region
where it is also very temperature-dependent. The most important
feature of the data is that the large negative drop present below
40 K in the zero-field Ce-115 data is progressively diminished as
the applied field is increased. While a field of 90 kOe does not
eliminate the drop in $R_{H}$ that occurs below 40 K, the
magnitude of the effect is diminished to the point where the Hall
response in the Ce-115 compounds is comparable to that seen in the
La-115 non-magnetic analogs. Increasing the field above roughly 5
kOe also produces a shallow minimum in $R_{H}$ that is centered
between 5 and 10 K. Below 15 K the $R_{H}$ field dependence
decreases monotonically with increasing field: changing the field
from the zero-field limit to 10 kOe reduces $R_{H}$ by roughly
50\%, while the Hall response in 50 and 90 kOe differ by less than
10\% relative to R$_{H}(H \rightarrow$ 0). The extreme
field-dependence exhibited by the Ce-115 compounds is in stark
contrast to the field-independent Hall response exhibited by the
La-115 compounds in the same temperature range. Measurements on
CeRhIn$_{5}$, CeIrIn$_{5}$, and CeCoIn$_{5}$ at 50, 100, and 300 K
in fields between 1 and 90 kOe indicate that the Hall response
becomes field-independent at these elevated temperatures.

The field-dependent Hall data shown in Fig. \ref{fig:HallH}
satisfy a scaling relationship that is similar to one used in
analyzing single-impurity magnetoresistance
data.\cite{Schlottmann89,Andraka94,Andraka95,Pietri00,Yamauchi00,Kaczorowski00,Christianson02}
Following the standard definition of the relative
magnetoresistance, {$\Delta\rho(H)/\rho(H=0)$}, we define the
relative change in the Hall coefficient as
\begin{equation}
\Gamma _{Hall} (H) = \frac{{R_H (H) - R_H (H \rightarrow 0)}}{{R_H
(H \rightarrow 0)}}. \label{eqn:gamma}
\end{equation}
Isotherms of the field-dependent Hall response can be superimposed
by plotting $\Gamma_{Hall}$ as a function of the transformed field
parameter $H^{*}$ defined by
\begin{equation}
H^*  = \frac{H}{\left( {T + T_o } \right)^\beta }
\label{eqn:Hscale}.
\end{equation}
The insets in Fig. \ref{fig:HallH} show $\Gamma_{Hall}$ plotted vs
$H^{*}$ for CeRhIn$_{5}$ (a), CeIrIn$_{5}$ (b), and CeCoIn$_{5}$
(c). In constructing these plots the scaling parameter $\beta$ was
set to 2, and the values used for the thermal scaling parameter
$T_{o}$ are listed in the insets. Each inset contains ten
superimposed isotherms that were measured at temperatures ranging
from 4 to 25 K (8 to 30 K for CeCoIn$_{5}$), and each isotherm is
composed of Hall data that was measured in fields ranging from 1
to 90 kOe. Scaling works best for $\beta$ = 2 $\pm$ 0.1, and the
$T_{o}$ values can be varied by $\pm$ 0.3 K without adversely
effecting the analysis; attempts to use the spin-1/2 MR exponent
($\beta$ = 1)\cite{Schlottmann89} were unsuccessful regardless of
the value used for $T_{o}$. Qualitatively, $\Gamma_{Hall}$ varies
with $H^{*}$ in roughly the same manner for all three Ce-115
compounds. Quantitatively, the CeCoIn$_{5}$ data are a stronger
function of $H^{*}$ than are the CeRhIn$_{5}$ and CeIrIn$_{5}$
data.

\begin{figure}
 \includegraphics{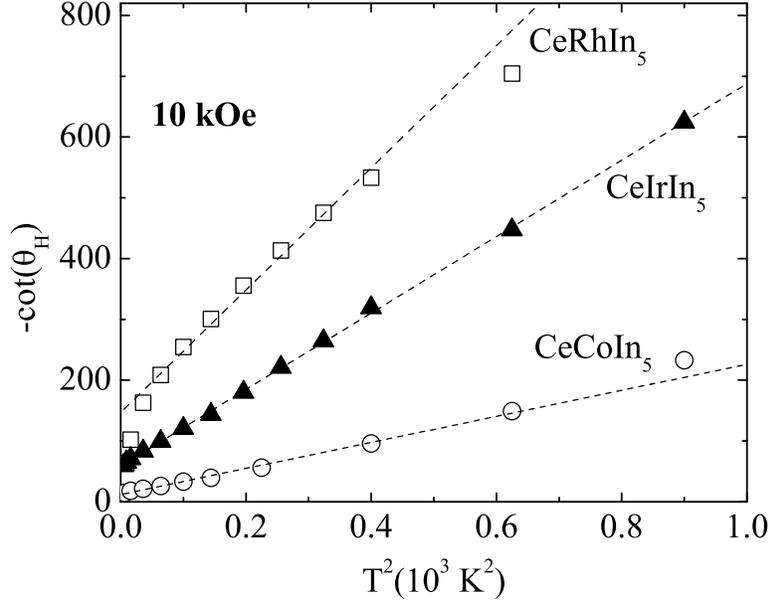}
 \caption{\label{fig:cotCe}In-plane Hall angle $\cot(\theta_{H}$) plotted as a function of $T^{2}$ for CeRhIn$_{5}$,
  CeIrIn$_{5}$, and CeCoIn$_{5}$. The dashed lines are
 linear fits to the data. The CeIrIn$_{5}$ data has been vertically offset by 50 for clarity.}
\end{figure}

We turn now to the Ce-115 Hall angle, which is plotted as
$-\cot(\theta_{H})$ vs $T^{2}$ in Fig. \ref{fig:cotCe}. The data
fall on straight lines for temperatures less than 30 K, indicating
that $\cot(\theta_{H})$ varies with temperature in a manner
consistent with Eq. (\ref{eqn:LaCot}). While 10 kOe Hall data were
used in constructing Fig. \ref{fig:cotCe}, essentially the same
temperature dependence results if data at other fields are used;
all that changes is the overall magnitude of $\cot(\theta_{H})$.
The temperature range over which the data follow a quadratic
temperature dependence are as follows: 8 to 20 K for CeRhIn$_{5}$,
4 to 30 K for CeIrIn$_{5}$, and 3 to 25 K for CeCoIn$_{5}$.
$\cot(\theta_{H})$ is nearly constant at higher temperatures
because $R_{H}$ and $\rho_{xx}$ both become weakly T-dependent
above 50 K. As with the La-115 compounds, the Ce-115 data vary
quadratically with temperature in the same temperature range where
$R_{H}$ exhibits considerably temperature dependence. The one
important difference between the Ce and La Hall angle data
concerns the temperature range over which $\cot(\theta_{H})$
varies quadratically with temperature. For the La-115 compounds
quadratic behavior is evident from 30 to 100 K, while the Ce-115
compounds show this behavior over a more limited temperature
range.

\section{\label{discuss}Discussion}

The Ce-115 Hall response is very different from what is observed
in most Ce Kondo-lattice systems. The canonical Ce heavy-fermion
Hall effect is dominated by a positive skew-scattering
contribution that dwarfs the conventional charge-carrier
contribution.\cite{Fert87} In contrast, the Hall effect in the
Ce-115 compounds appears to be governed by the conventional
charge-carrier contribution. After disentangling the relative
contributions from these two mechanisms it will become clear that
Kondo interactions and $f$-electron effects play an important part
in determining the temperature and field dependence exhibited by
the Hall effect in the 115 compounds.

The Hall response in heavy-fermion compounds is produced by a
contribution from skew-scattering $R_{H}^{skew}$ and the ordinary
Hall effect $R_{H}^{o}$,
\begin{equation}
R_H  = R_H^{skew}  + R_H^o \label{eqn:RHtot}.
\end{equation}
The skew scattering term stems from interactions between the large
Ce spin state and the applied field that produce a left-right
asymmetry in charge-carrier scattering. As first expressed by Fert
and Levy,\cite{Fert87} the skew-scattering terms takes the form
\begin{equation}
R_H^{skew}  = \xi \rho _{mag} \mathop \chi \limits^ \sim
\label{eqn:RHskew},
\end{equation}
where $\tilde{\chi}$ = $\chi/C$ is the reduced magnetic
susceptibility and $C$ is the Curie constant. For $T \gg T_{K}$
the parameter $\xi$ becomes
\begin{equation}
\xi  =  - \frac{5}{7}g\frac{{\mu _B }}{{k_B }}\sin \delta \cos
\delta \label{eqn:RHxi},
\end{equation}
where $\delta$ is phase-shift produced by incoherent Kondo
scattering, $g$ is the magnetic ion's Land$\acute{e}$ g-factor,
$\mu_{B}$ is the Bohr magneton, and k$_{B}$ is Boltzmann's
constant. As the system is cooled below $T_{coh}$ incoherent Kondo
scattering dies off; calculations based on the periodic Anderson
Hamiltonian suggest that $R_{H} \sim \rho_{mag}^{2}$ in the
coherent regime.\cite{CohHall} These theoretical results indicate
that, starting from $T$ = 0, $R_{H}^{skew}$ should increase
rapidly from zero, achieve a broad maximum at the temperature
where $\rho_{mag}$ peaks, and gradually decrease at higher
temperatures. The Hall effect in many Ce, U, and Yb heavy-electron
systems behave in this
manner.\cite{Cattaneo85,Penney86,Winzer86,Fert87,Hadzic86,Hiraoka92,Hundley01}
These compounds exhibit a positive skew-scattering Hall response
because the repulsive single-impurity scattering potential leads
to a negative phase shift.

\begin{figure}
 \includegraphics{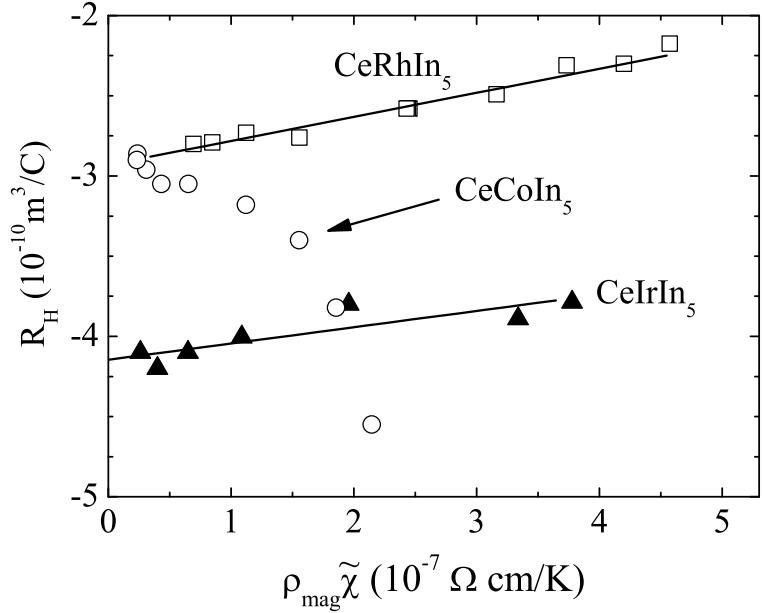}
 \caption{\label{fig:skew}R$_{H}$ plotted versus $\rho_{mag}\widetilde{\chi}$ for CeRhIn$_{5}$, CeIrIn$_{5}$, and CeCoIn$_{5}$.
 Temperature is an implicit parameter in this figure; 300 K data appear on the left side of the
 plot and 40 K data appear to the right. The solid lines are linear fits to the Rh and Ir data.}
\end{figure}

Careful analysis of the Hall data indicates that skew-scattering
is a minor contributor to the overall Hall response in the Ce-115
compounds. This is particularly true below 50 K where the large
negative drop in the Hall coefficient is inconsistent with
skew-scattering. The most obvious inconsistency is that the low-T
anomaly occurs below the coherence temperature where any skew
scattering contribution should be dropping to zero. Additionally,
the sign of the Hall feature is inconsistent with skew scattering,
$R_{H}$ does not peak at $T_{coh}$, and the sharp drop in the Hall
response cannot be fit to $\rho_{mag}\tilde{\chi}$. Skew
scattering only becomes a significant contributor to the Hall
effect above 50 K. This is shown in Fig. \ref{fig:skew} where
$R_{H}(T)$ is plotted vs $\rho_{mag}\tilde{\chi}$ for temperatures
from 40 to 300K. The CeRhIn$_{5}$ and CeIrIn$_{5}$ Hall data vary
linearly with $\rho_{mag}\tilde{\chi}$, as predicted by Eq.
(\ref{eqn:RHskew}). The parameters determined from the figure are
$\xi$ = 0.016 K/T and $R_{H}^{o}$ = -2.9 x 10$^{-10}$ m$^{3}$/C
for CeRhIn$_{5}$, and $\xi$ = 0.010 K/T and $R_{H}^{o}$ = -4.1 x
10$^{-10}$ m$^{3}$/C for CeIrIn$_{5}$. The resulting phase-shifts
($\delta_{Rh}$ = -0.042 radians and $\delta_{Ir}$ = -0.025
radians) are consistent with those reported for other Ce
heavy-fermion systems.\cite{Fert87,Hadzic86,Hundley01} For
CeCoIn$_{5}$ $R_{H}$ does not scale with $\rho_{mag}\tilde{\chi}$.
This suggests that any skew-scattering contribution present in the
Co material is overwhelmed by the conventional Hall term. The
skew-scattering contribution evident in the Ce-115 Hall data is
small, in part, because $\rho_{mag}$ is five to ten smaller than
in other heavy-fermion systems.

Similarities in the Hall response of the Ce and La 115 compounds
below 100 K imply that the second term in Eq. (\ref{eqn:RHtot}) is
responsible for the anomalous drop present in the Ce-115 data.
dHvA measurements\cite{Shishido02} indicate that the La and Ce
compounds share essentially the same electronic structure, so the
conventional Ce-115 Hall term should mimic that of the La-115
materials. Band-structure calculations and dHvA measurements
indicate that the 115's are compensated materials ($n_{e} =
n_{h}$) with multiple electron and hole Fermi surfaces that form
complex 2D and 3D structures.\cite{Haga01,Hall01a,Shishido02} The
Hall effect of a multiband system is determined by the weighted
sum of the contributions from each band. Qualitatively, a
temperature-dependent Hall coefficient can occur when multiple
electron and hole band cross the Fermi energy and the bands have
mobilities with different temperature dependencies.\cite{HurdHall}
The situation for the 115's is even more complicated since the
electron and hole conduction bands give rise to highly anisotropic
Fermi surfaces, and, presumably, anisotropic relaxation times.
This description also applies to the electronic structures of
MgB$_{2}$ and YBa$_{2}$Cu$_{3}$O$_{7}$, both of which also have a
Hall coefficient described by Eq.
(\ref{eqn:LaHall}).\cite{Chien91,Jin01} While the presence of a
complex FS qualitatively accounts for the temperature-dependent
La-115 Hall coefficient, it does not address the question of why
$R_{H}$ becomes increasingly negative below 100 K. The
electron-like Hall response may occur because the electron
extremal orbits seen in LaRhIn$_{5}$ dHvA spectra tend to have
lighter masses, and hence larger mobilities, than the extremal
hole orbits.\cite{Shishido02}

\begin{figure}
 \includegraphics{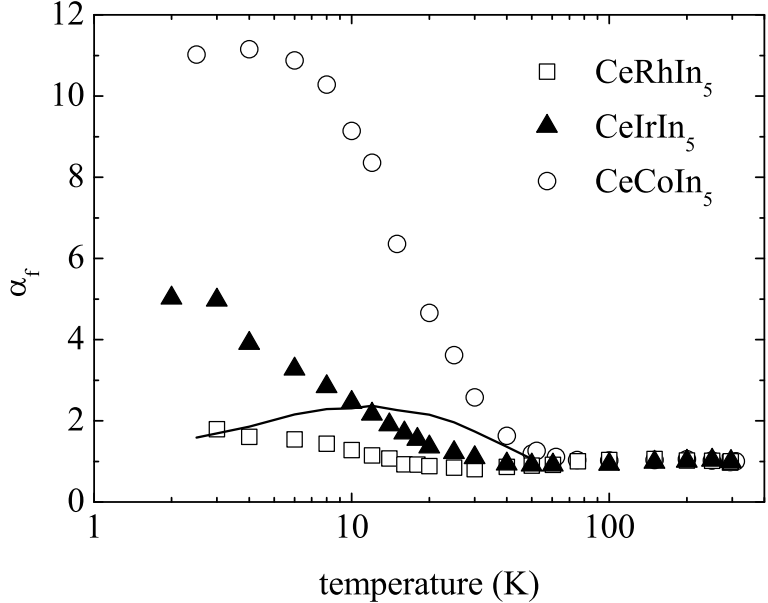}
 \caption{\label{fig:Hallf}The low-field $f$-electron Hall weighting function $\alpha_{f}$
 plotted as a function of temperature for CeRhIn$_{5}$,
 CeIrIn$_{5}$, and CeCoIn$_{5}$. The solid line shows $\alpha_{f}$ in a field of 90 kOe for CeCoIn$_{5}$.}
\end{figure}

Despite their similarities, there are also significant difference
in the Ce-115 and La-115 Hall response. The most obvious
difference is that the Hall anomalies are much larger in the
Ce-115 compounds. The Hall response of CeCoIn$_{5}$ changes by a
factor of 20 between 100 and 4 K; the Hall coefficients of
CeRhIn$_{5}$ and CeIrIn$_{5}$ change by a factor of 4 and 10,
respectively, over the same temperature range. In comparison, the
Hall response at 100 K and 4 K differ by only a factor of two in
the La-115 compounds. These differences presumably stem from the
influence of the $f$-electrons that are present in the Ce
compounds. If we assume that the Ce-115 Hall response
$R_{H}^{(Ce)}$ is the sum of a skew-scattering term and a term
proportional to the La-115 coefficient, $R_{H}^{(Ce)}$ can be
expressed as
\begin{equation}
R_H^{(Ce)} (H,T) = R_H^{skew} (T) + \alpha _f (H,T)R_H^{(La)} (T),
\label{RH:Ce}
\end{equation}
where $\alpha_{f}$ is defined as the $f$-electron Hall weighting
function. Figure \ref{fig:Hallf} shows $\alpha_{f}(H \rightarrow
0)$ (symbols) plotted vs temperature for the three Ce-115
compounds; the solid line in the figure shows $\alpha_{f}$(90 kOe)
for CeCoIn$_{5}$. $\alpha_{f}$ was determined from Eq.
(\ref{RH:Ce}) by combining Ce and La Hall data with the skew
scattering contribution determined from the analysis associated
with Fig. \ref{fig:skew} (we assume $R_{H}^{skew}$ = 0 for
CeCoIn$_{5}$). $\alpha_{f}$ is unity from roughly 50 to 300 K,
indicating that the conventional Hall contribution in the Ce
compounds matches what is measured for the La-115's. Below 50 K
$\alpha_{f}$ monotonically increases with decreasing temperature,
and saturates below 3 K; the rise in $\alpha_{f}$ begins at 45 K
for CeCoIn$_{5}$ and 25 K for both CeRhIn$_{5}$ and CeIrIn$_{5}$.
These onset temperatures indicate that the $f$-electron
contribution to the Hall effect begins to grow at roughly the same
temperature where the resistivity peaks. As such, the growth in
$\alpha_{f}$ evident below 50 K appears to be correlated with the
commencement of Kondo coherence.

The extremal orbit masses observed in dHvA spectra can provide a
simple explanation for the connection between the Ce-115 Hall
response and Kondo interactions below 50 K. Those
measurements\cite{Settai01} on CeCoIn$_{5}$ detect a heavy hole
orbit with a mass (87$m_{o}$) that is consistent with the large
electronic specific-heat coefficient observed
experimentally.\cite{Petrovic01b} In comparison, the CeCoIn$_{5}$
electron orbits are significantly lighter ($\sim$15$m_{o}$). This
extreme electron-hole mass asymmetry would lead to a mobility
asymmetry, and, ultimately, an even larger negative Hall anomaly
than is seen in LaCoIn$_{5}$. Although the differences are not as
extreme, CeIrIn$_{5}$ and CeRhIn$_{5}$ also exhibit asymmetries in
their respective hole and electron
masses.\cite{Haga01,Hall01a,Shishido02} Of the three Ce-115
compounds, CeRhIn$_{5}$ has the smallest electron-hole mass
asymmetry, and the least amount of 4f character in its Fermi
surface;\cite{Cornelius00,Shishido02} these FS features are
consistent with the fact that CeRhIn$_{5}$ also has the smallest
low-T Hall anomaly in the Ce-115 series. The enhanced low-T Ce-115
carrier masses are a direct result of the Kondo interactions that
produce the large electronic contribution to $C_{p}(T)$. The Kondo
resonance gradually develops with decreasing temperature, so that
the carrier masses will not be heavy for $T \gg T_{K}$. The
temperature-dependence shown by the $f$-electron Hall weighting
function is then simply a reflection of the carrier mass
enhancement that gradually develops as the system is cooled below
$T_{coh}$.

The substantial field dependence present in the Ce-115 Hall
response offers confirming evidence that field-dependent many-body
interactions are responsible for the sizable difference between
the Ce and La low-T Hall anomalies. An applied magnetic field has
a deleterious effect on the heavy-fermion state because it tends
to broaden the Kondo resonance and shift it below the Fermi
energy.\cite{Satoh85,Lorek91,Edwards97} The $f$-electron
contribution to the FS in-turn drops, and the Sommerfeld
coefficient and the large zero-field effective mass are
reduced.\cite{Satoh85,Andraka91} The solid line in Fig.
\ref{fig:Hallf} shows that the $f$-electron contribution to the
CeCoIn$_{5}$ Hall effect is substantially reduced in 90 kOe; the
same is true for CeRhIn$_{5}$ and CeIrIn$_{5}$. These results are
consistent with the large field-induced reduction of the
CeCoIn$_{5}$ carrier mass observed in dHvA
measurements.\cite{Settai01} The analysis used in Sec. \ref{Ce115}
to parameterize $R_{H}$(H,T) data also shows a link between the
Ce-115 Hall response and Kondo interactions. The H-T scaling
analysis used on the Ce-115 Hall data is similar to the
parametrization that can be applied to single-impurity MR
data.\cite{Schlottmann89} The MR of a spin-1/2 Kondo system
typically follows the H-T scaling expressed by Eq.
(\ref{eqn:Hscale}) with $\beta$ = 1, and $T_{o}$ = T$_{K}$. The
Ce-115 $T_{o}$ values listed in Fig. \ref{fig:Hallf} are roughly
consistent with the Kondo temperatures estimated from low-T
Sommerfeld coefficients.\cite{Hegger00,Petrovic01a,Petrovic01b}.
Hence, the $R_{H}^{(Ce)}(H,T)$ data and $\alpha_{f}(H,T)$ values
are consistent with a field-induced suppression in the
$f$-electron character of the Ce-115 Fermi-surface states;
further, H-T Hall data scaling is consistent with Kondo energy
scales of a few degrees Kelvin.

Our analysis bears some resemblance to the two-fluid Kondo lattice
model proposed by Nakatsuji, \textit{et al.}\cite{Nakatsuji04}
This model divides a Kondo lattice system into a Kondo-gas
component (analogous to a Kondo-impurity phase) and a Kondo-liquid
component (analogous to a coherent heavy-Fermion phase), with the
temperature-dependent evolution of Kondo-lattice properties
controlled by the mixing parameter $f(T)$. The gas phase,
characterized by a single-ion Kondo scale $T_{K}$, dominates the
system's properties at high temperatures. The liquid phase,
characterized by the intersite coupling energy scale $T^{*}$,
begins to influence the system's physical properties at
temperatures less than $T^{*}$. A two-fluid
analysis\cite{Nakatsuji02,Nakatsuji04} of $\chi$(T) and $C_{p}(T)$
data finds that the energy scales in CeCoIn$_{5}$ are $T_{K}$ =
1.7 K and $T^{*}$ = 45 K. The model's crossover from local-moment
behavior at high temperatures to itinerant heavy-fermion behavior
at low temperatures, and the energy scales derived for
CeCoIn$_{5}$, accurately describe the Ce-115 $R_{H}(T)$ data. It
is particularly noteworthy that the increasing importance of the
heavy Kondo-liquid phase below $T^{*}$ in the two-fluid model
provides a simple explanation for the rise in $\alpha_{f}(T)$ that
occurs in the Ce-115 Hall data below 50 K.

Lastly, we consider the significance of the $\cot(\theta_{H}) \sim
T^{2}$ behavior present in the 115 Hall data. The same
T-dependence is present in high-T$_{c}$ cuprate
data\cite{Chien91,Ginsberg93} and some have suggested that this is
linked\cite{Kontani99} to a QCP. The NFL behavior evident in the
physical properties of CeCoIn$_{5}$ (and possibly CeIrIn$_{5}$)
below $\sim$ 5 K might also be related to a QCP. As such, it is
tempting to ascribe the novel $\cot(\theta_{H})$ temperature
dependence in these two Ce-115 materials to critical spin
fluctuations associated with a nearby QCP. This interpretation
appears untenable since the Hall angle data of CeRhIn$_{5}$ and
the three La-115 compounds -- all of which show no QCP-related
phenomena -- also exhibit the $\cot(\theta_{H})\sim T^{2}$
behavior. A more credible conclusion is that the quadratic
Hall-angle temperature-dependence results from the peculiar
``conventional'' Hall response intrinsic to the 115 electronic
structure.

\section{\label{conclude}Conclusions}

The Hall response of the Ce-115 compounds differs markedly from
that of most Kondo systems. $R_{H}^{(Ce)}$ is dominated by the
conventional Hall effect rather than that due to skew-scattering.
This comes about because of the complex electronic structure
intrinsic to the 115 system. The field and temperature-dependent
variation of the Ce-115 Hall coefficients below 50 K are
consistent with the same Kondo interactions that also influence
other transport and thermodynamic properties. These results
indicate that conventional transport mechanisms cannot always be
ignored in interpreting the physical properties of $f$-electron
systems.

\begin{acknowledgments}
 We thank E. D. Bauer, J. M. Lawrence, and Z. Fisk for encouragement and useful
 discussions. We also thank N. O. Moreno for assistance with PPMS
 Hall measurements. Work at Los Alamos was performed under the
 auspices of the US Department of Energy.
\end{acknowledgments}


\end{document}